\documentclass[aps,prl,twocolumn,showpacs,superscriptaddress]{revtex4}
\usepackage{graphicx}
\begin{document}


\title{\bf Superconductivity:  Exotic Commonalities in Phase and Mode\/}  
     \author{Y.~J.~Uemura}
     \affiliation{Department of Physics, Columbia University, New York, New York 10027, USA}
\date{Feb. 8, 2009}

 

\maketitle

{\bf
Recent muon and neutron experiments on the new FeAs-based superconductors 
revealed phase diagrams characterized by first-order evolution from antiferromagnetic
to superconducting states, and an inelastic magnetic resonance mode whose
energy scales as $\sim 4 k_{B}T_{c}$.  These features exhibit striking
commonalities with cuprate, backyball, organic, and heavy-fermion superconductors as well as
superfluid $^{4}$He.\/}\\

For every new superconducting system, determination of the phase diagram is both a starting
point and a major goal of experimental studies.  Muon spin relaxation ($\mu$SR)
and neutron scattering are two strong particle probes for this purpose.
In the present issue, two muon groups report their findings in 
RE(O,F)FeAs with RE=La \cite{luetkens} and Sm \cite{drew}.
Together with the earlier neutron study on RE=Ce \cite{zhao}, these represent first sets
of the phase diagrams of the ``1111'' FeAs superconductor family first discovered in 
February of 2008 \cite{hosono}.  As seen in Fig. 1(a), the three results exhibit differences in
details, but they indicate that the superconducting (SC) state takes over the
antiferromagnetic (AF) state, with abrupt disappearance of the AF state for RE=La,
and phase-separated coexistence of AF and SC regions in RE=Sm. 
 
As compared to neutron scattering, $\mu$SR has 
better sensitivities to static magnetic order
with random or small moments, and can clearly determine the volume fraction of 
magnetically ordered regions near the phase boundaries \cite{mnsi113}. These advantages
are fully employed in the studies of the La and Sm systems.
A more recent muon study in the RE=Ce system
near the SC-AF boundary \cite{uemura} indicates that the RE=Ce system also
shows evolution similar to that of RE=La.
All together, we observe behavior significantly different from the standard picture of
second-order evolution from the AF to SC states associated with a quantum critical point (QCP).
The abrupt disappearance of the magnetic phase and phase separation of the AF and SC
states are both characteristic of first-order quantum evolution at $T \rightarrow 0$.

As shown in Fig. 1(b)-(e), these features are found in 
various other superconductors, including the ``122'' FeAs system
(Ba,K)Fe$_{2}$As$_{2}$ 
by neutrons \cite{bak122neutron} (1(b)) and muons \cite{gokocondmat},
the YBCO cuprate system (1(c)) by muons \cite{sannaybco}, alkali-doped 
A$_{3}$C$_{60}$ (1(d)) by muons and EPR \cite{a3c60}, 
organic BEDT by NMR \cite{bedt}, CeRhIn$_{5}$ (1(e)) by transport and calorimetry \cite{cerhin5},
and CeCu$_{2}$Si$_{2}$ by muons \cite{lukeprl}.
The boundary of the SC and static spin stripe states in the 214 cuprate systems is
associated with phase separation \cite{214}.  
The phase diagram for superfluid $^{4}$He also looks quite similar (Fig. 1(f)).

Spectacular commonality also exists with the energy scale of neutron 
inelastic excitations observed in the 
SC (or superfluid) state.  Recent observation of the magnetic resonance mode in 
(Ba,K)Fe$_{2}$As$_{2}$ \cite{resonance122} (Fig. 2(a)) follows earlier results 
in the cuprates \cite{bourges}, CeCoIn$_{5}$ \cite{broholmcecoin5}, and
CeCu$_{2}$Si$_{2}$ \cite{sces2007}.  The magnetic resonance mode appears with 
the same symmetry as the AF correlations (shown by the blue closed circle in Fig. 2(a)),
as short-range and dynamic spin correlations related to the AF state.
In this sense, the resonance mode may be analogous to rotons in superfluid $^{4}$He
(Fig. 2(b) \cite{he}), which are inelastic soft phonon modes associated with the 
imminent HCP solid state whose Bragg points are shown by the blue closed circles in Fig. 2(b).

Indeed, when compared in a plot of $T_{c}$ versus mode energy $\hbar\omega$ in Fig. 2(c)
\cite{uemura,yamazakiprize,he}, all these excitations exhibit nearly identical linear slopes of 
$\hbar\omega \sim 4 k_{B}T_{c}$.  The resonance mode in cuprates appears with an
``hour-glass''-shaped dispersion \cite{cuprate}.  
If we use the energy of the lower end of populated states in this dispersion curve,
often referred to as the spin-gap energy and shown by closed squares in Fig. 2(c), the slope agrees
well with those of all the other systems.  The energy of the inelastic A$_{1g}$ Raman mode 
in cuprates \cite{sacuto} also
follows this correlation, hinting that not only a spin phenomenon but also a charge phenomenon might join this argument
\cite{yamazakiprize}.

Figure 2(c) suggests that the transition temperature $T_{c}$ of the SC (or superfluid)
state may be determined by
the energy of such ``soft-mode'' dynamic excitations related to the adjacent
antiferromagnetic (or solid) state, whose energy represents the ``closeness to the 
competing AF (or HCP) state''.
The first-order-like evolution, seen in Fig. 1, actually helps existence of 
dynamic modes which are often very sharp in both momentum and energy space.  
In systems exhibiting second
order transitions, at points somewhat away from the QCP, one would expect
much broader and diffusive inelastic responses.
Although the details need to be clarified by further studies, these exotic commonalities
suggest the importance of strong coupling and collective modes, and promote
researchers to think beyond textbook cases of 
second order transitions, quantum critical points,
weak coupling, and single (quasi) particle descriptions.

The author acknowledges financial support by the NSF MWN-CIAM program
DMR 05-02706 and 08-06846.

\onecolumngrid
\vfill \eject
\newpage
\begin{figure}[h]

\begin{center}
\vskip 1.0 truecm
\includegraphics[angle=0,width=6.0in]{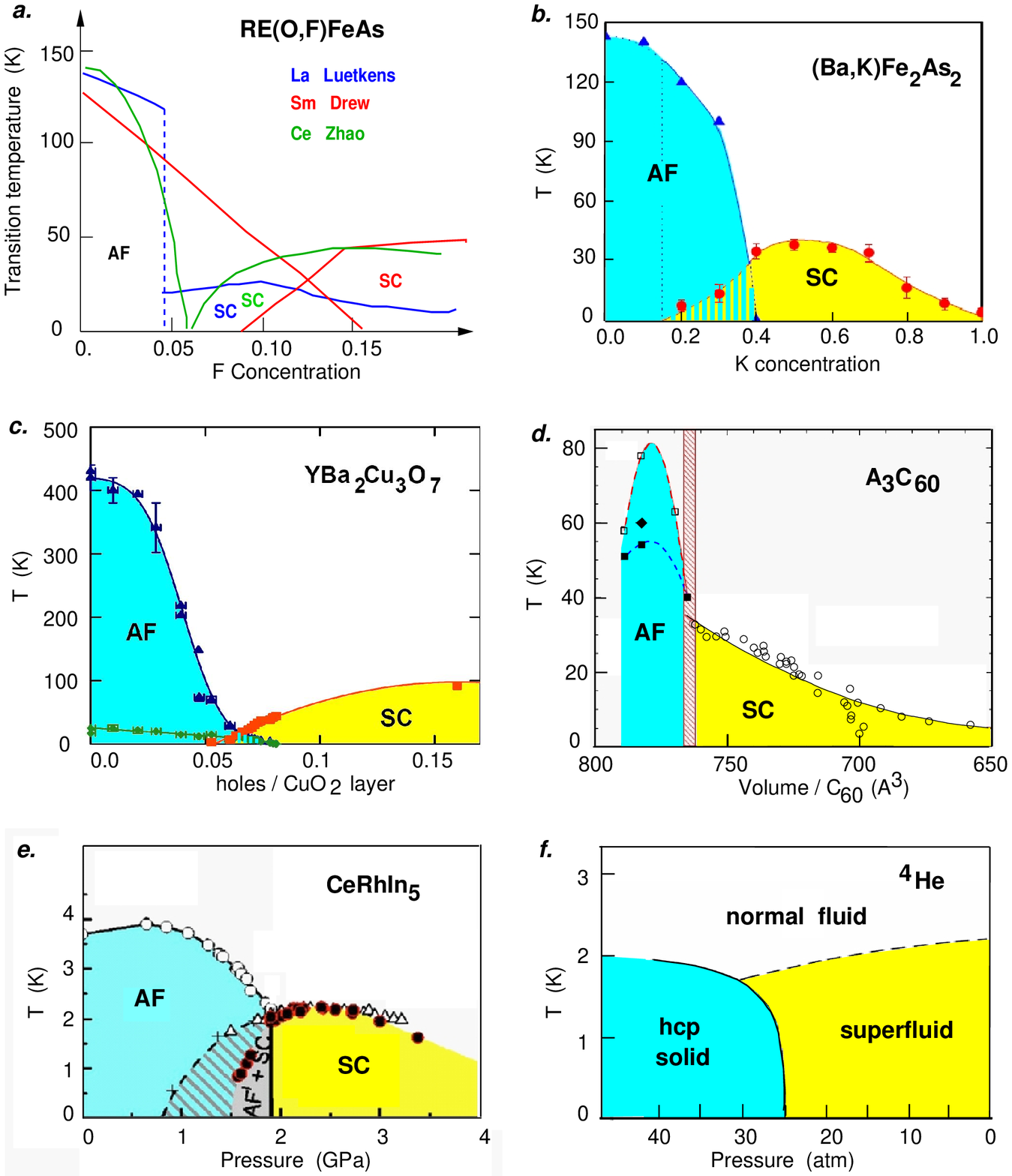}
\vskip 1.0 truecm
\label{Figure 1.} 

\caption{\label{Figure 1.}
(color) Electronic phase diagram, as functions of composition, pressure, and/or unit-cell volume in (a) RE(O,F)FeAs (RE = La, Sm, Ce)\cite{luetkens,drew,zhao},
(b) (Ba,K)Fe$_{2}$As$_{2}$ \cite{bak122neutron}, 
(c) YBa$_{2}$Cu$_{3}$O$_{7-\delta}$ \cite{sannaybco},
(d) A$_{3}$C$_{60}$ (A = K, Cs, Rb) \cite{a3c60}, and
(e) CeRhIn$_{5}$ \cite{cerhin5}  
(CeCoIn$_{5}$ at ambient pressure corresponds to CeRhIn$_{5}$ at $p \sim 2.4$ GPa).    
Figure 1(f) shows the phase diagram of
superfluid $^{4}$He.  All these systems exhibit abrupt disappearance of
the antiferromagnetic (AF) or HCP solid phase or coexistence of 
the AF and the superconducting (SC) phases near the phase boundary.}
\end{center}
\end{figure}
\vfill \eject
\newpage

\begin{figure}[t]

\begin{center}

\includegraphics[angle=0,width=6.5in]{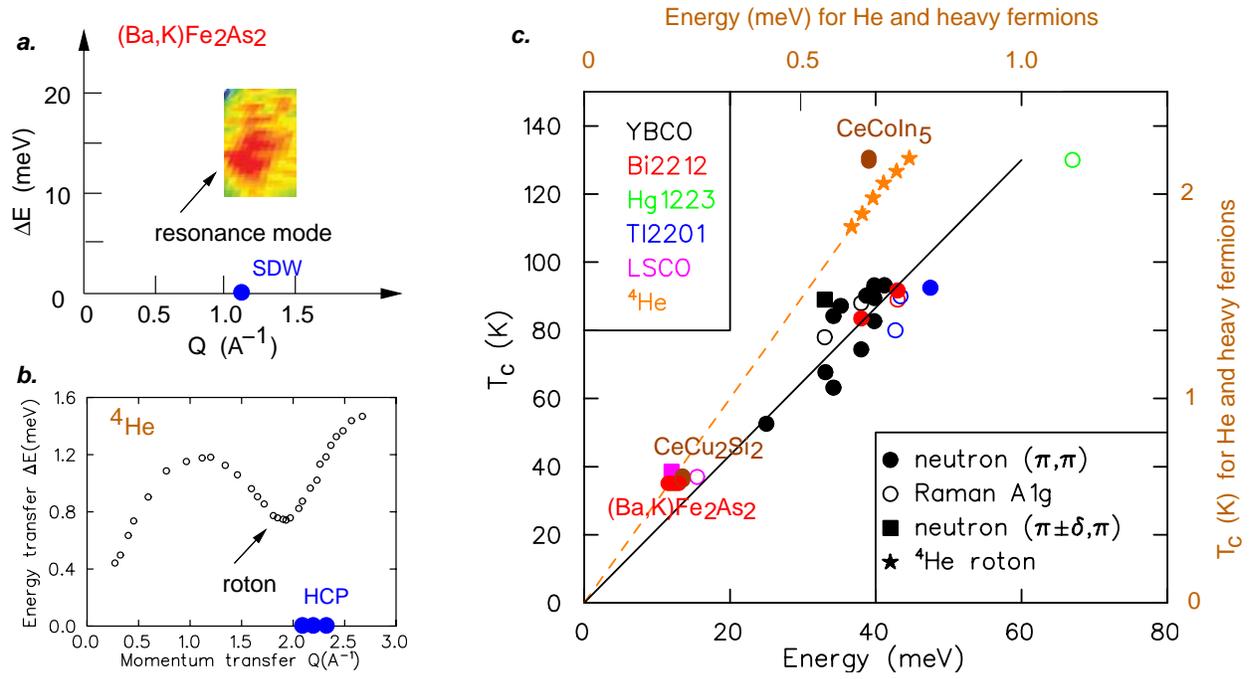}
\vskip 1.0 truecm 

\label{Figure 2.} 

\caption{\label{Figure 2.} 
(color) (a) Schematic view of the magnetic resonance mode
in (Ba,K)Fe$_{2}$As$_{2}$ \cite{resonance122}, with the closed blue
circle denoting the momentum transfer of the 2-dimensional (1/2,1/2,0) 
antiferromagnetic correlations.  (b) Phonon-roton dispersion relation 
in superfluid $^{4}$He, with the closed blue circles denoting
the Bragg points of the hexagonal closed-packed (HCP) phase of solid He \cite{he}.
(c) Correlations between the transition temperature $T_{c}$ and 
the energy $\hbar\omega$ of the magnetic resonance mode observed in 
the superconducting state of the  
high-$T_{c}$ cuprate systems \cite{bourges,cuprate},
(Ba,K)Fe$_{2}$As$_{2}$ \cite{resonance122},
CeCoIn$_{5}$ \cite{broholmcecoin5}, and
CeCu$_{2}$Si$_{2}$ \cite{sces2007}.  The closed square symbols denote
the ``spin-gap'' energy obtained from the low-energy end of the 
hour-glass dispersion shape \cite{cuprate}.  The star symbols represent
the lambda-point superfluid transition temperature $T_{c}$ and the roton 
energies in superfluid $^{4}$He at ambient and applied pressure \cite{he}.
The right-vertical and top-horizontal axes for He, CeCoIn$_{5}$ and CeCu$_{2}$Si$_{2}$ are  
both scaled by a factor 60 with respect to 
the left and bottom axes for the other systems.  The aspect ratio 
is preserved, however, for direct comparisons of the slope $T_{c}/\hbar\omega$
of all the different systems.  Updated after ref. \cite{yamazakiprize} and adopted from  
ref. \cite{uemura}.}
\end{center}
\end{figure}
\vfill \eject
\end{document}